\documentclass[11pt,twoside]{article}
\usepackage{asp2010}

\resetcounters

\begin{document}

\title{Helioseismology from SODISM and HMI Intensity Images}
\author{T.~Corbard,$^1$ D.~Salabert,$^{1,2}$ P.~Boumier,$^3$ and the PICARD team}
\affil{$^1$ Laboratoire Lagrange, UMR7293, 
 Universit\'e de Nice Sophia-Antipolis, CNRS, Observatoire de la C\^ote d'Azur,
 Bd. de l'Observatoire, 06304 Nice, France}
\affil{$^2$ CEA/IRFU/Service d'Astrophysique, AIM, CE Saclay, 91191 Gif sur Yvette, France}
\affil{$^3$ Institut d'Astrophysique Spatiale, CNRS-Universit\'e Paris XI, UMR 8617, 91405 Orsay Cedex,
 France}

\begin{abstract}
Continuum intensity images from PICARD/SODISM and SDO/HMI 
covering a 209-day period in 2011 are analyzed
in order to extract  mode parameters for spherical harmonics up to $l=100$.
SODISM helioseismology signal is affected by the low orbit of PICARD and by 
important gaps and CCD persistence effects. 
SODISM intensity signal has a lower signal to noise ratio and duty cycle than HMI and 
less modes were successfully fitted over the same period. A comparison of the rotation profiles
obtained from both sets of continuum images shows however that the results remain compatible 
within one standard deviation of HMI formal errors.

\end{abstract}

\section{Introduction}
 One-minute cadence low-resolution  images 
($256^2$ pixels of about $8\times8$ arc-seconds) recorded by the SODISM telescope  \citep{Meftah2013}
 on board PICARD satellite \citep{Thuillier2006} 
are available for a helioseismology ``medium-$l$'' program in intensity \citep{Corbard2008}.
We first discuss the calibration steps of the raw SODISM data and then we 
compare the internal rotation rates inferred by inverting the frequency splittings  
obtained from both SODISM and HMI continuum images (at 535.7 and 617.3~nm respectively)
over the same 209-day period in 2011.

\section{Characteristics of SODISM signal and data calibration}

\articlefigure[angle=-90,width=0.8\textwidth]{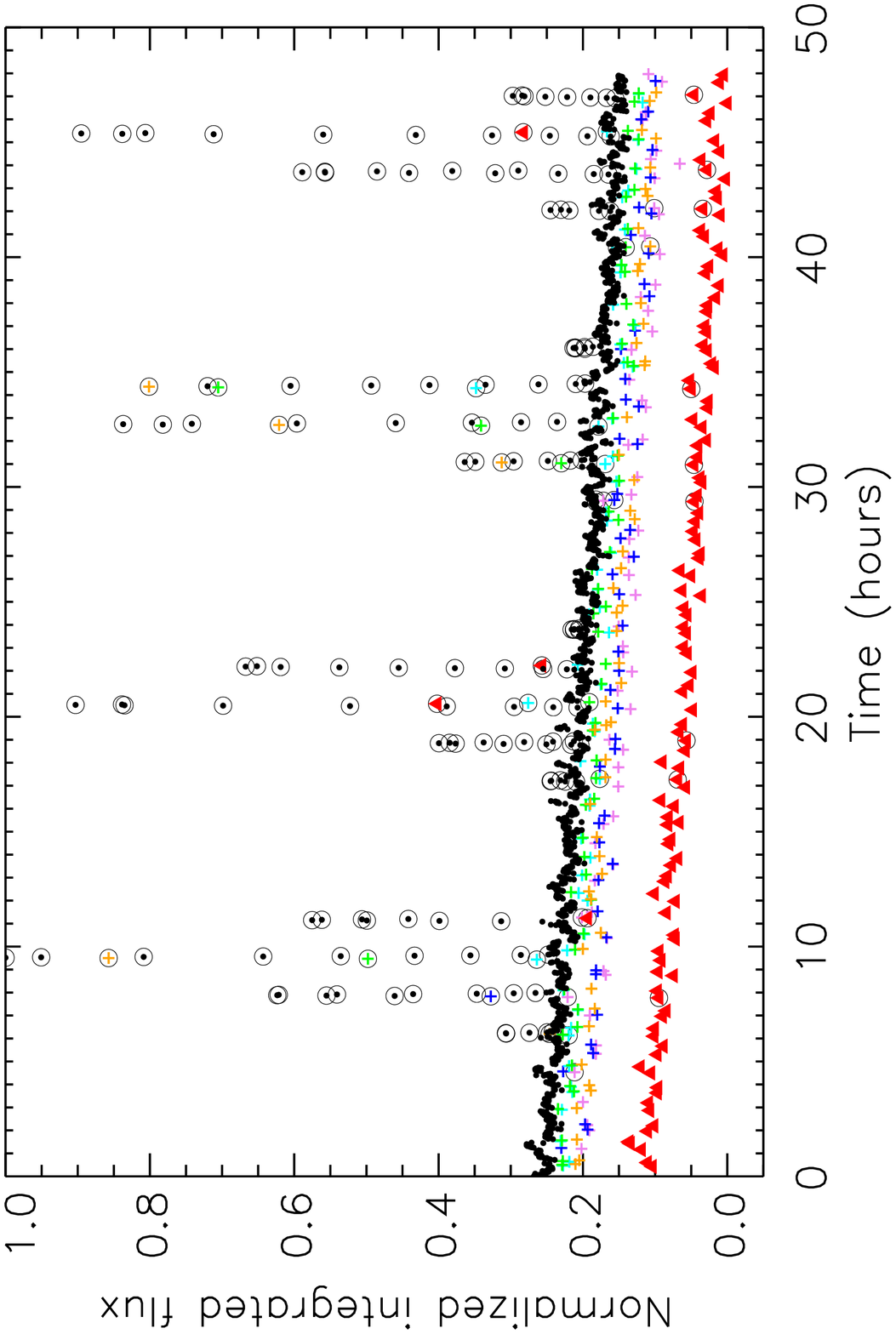}{Fig:rema}{Integrated intensity 
(i.e. $l=0$ signal)
 of all images made at 535~nm  for the helioseismology program during two consecutive days 
(May 16-17, 2011). 
 The plotting symbol used depends on what was recorded the minute
before: another image at 535~nm (dark dots), an image at another wavelength (crosses) or a dark
 current (triangles).
Circles indicate when the satellite was inside the SAA during image record.
The  color code for the crosses corresponds 
to the wavelength that was observed the minute before:
215~nm (violet); 393~nm (blue); 607~nm (green); 782~nm (orange) (see online color figure)}

Three main characteristics of the SODISM signal can be outlined:
\begin{itemize}
\item Regular passages through the South Atlantic Anomaly (SAA) due to the low orbit of the PICARD 
satellite \citep{Irbah2012}.
 On a 24-hr period, the SAA corresponds to about 7\% of the measurements which are affected 
(see Fig.~\ref{Fig:rema}).
\item Orbital period around the Earth of about 100 minutes produces aliased peaks in the power spectra of the low-degree p-mode signals.

\item Presence of CCD persistence affecting about 10\% of the images.
 \end{itemize}

Figure \ref{Fig:rema} clearly shows that the photometric level at a given minute depends on 
what was recorded 1-min before (dark current or images at different wavelengths for the astrometric
 program of PICARD).   Different photometric offsets are thus introduced in the
helioseismic observations. This is  the most problematic source of noise for
helioseismic analysis. As its origin is not fully understood yet \citep[see also][]{Hochedez2013}, a
posteriori ad hoc correction of the data is performed. 
The routine interruptions for the astrometric program also create
a 2-min aliasing in the power spectrum.

SODISM data calibration for helioseismic analysis must therefore include the following steps:
identification of the measurements taken in the SAA through the geo localization of the spacecraft,
ad hoc correction of the CCD persistence and high-pass filtering of the light curves by fitting e.g.
Legendre polynomials. Finally, gaps smaller than 5 minutes are filled using a linear prediction algorithm, 
which also helps in reducing the 2-min aliasing.

\articlefigure[width=0.6\textwidth]{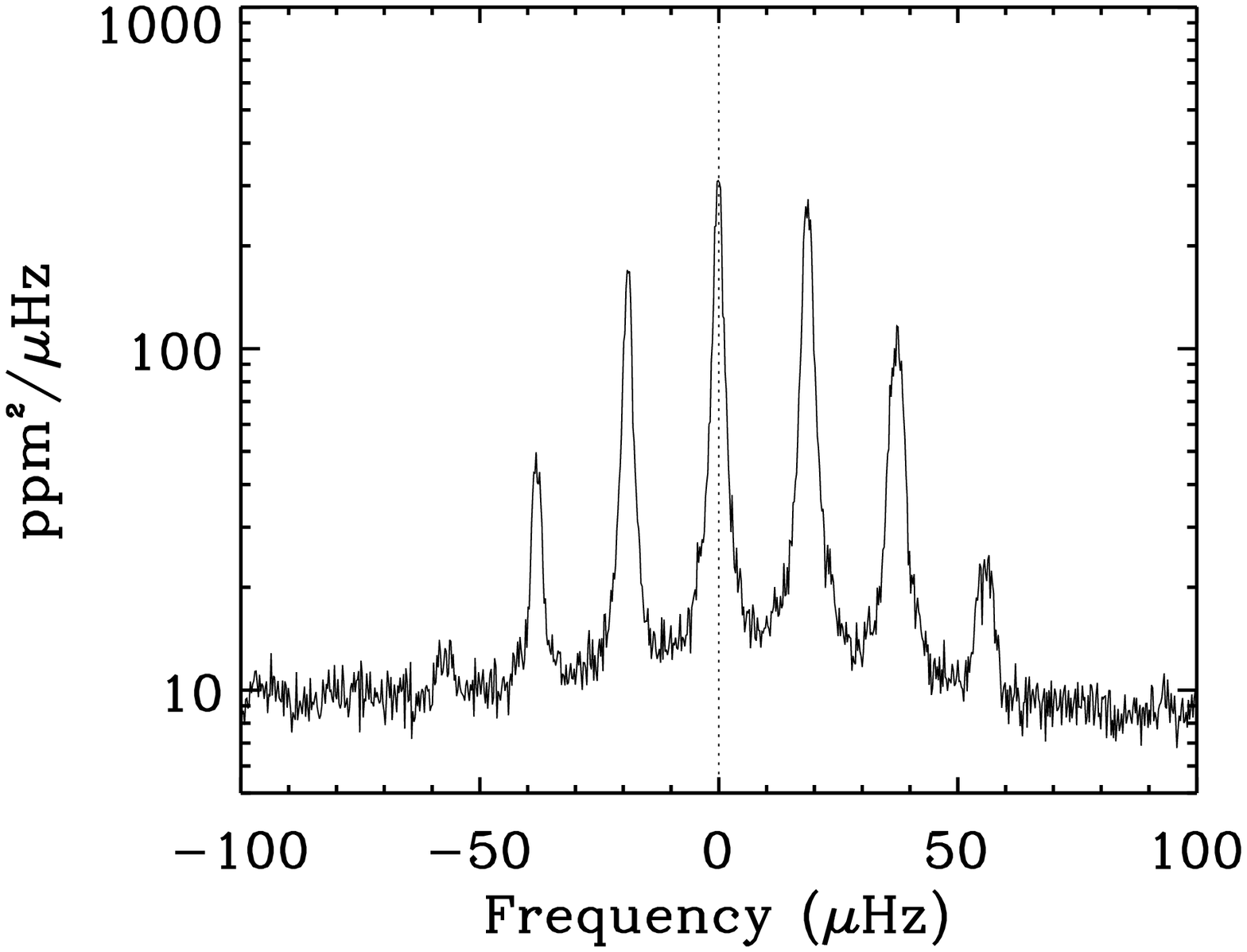}{Fig:assym}{Rotation-corrected, $m$-averaged spectrum
 of $l=50$, $n=10$ (indicated by vertical dotted line) in a 200 nHz frequency window.}

\section{Results and comparison with HMI}
The power spectra, mode parameters and internal rotation rates obtained from SODISM images at 535~nm 
for the 209-day  period going from 2011 April 16 to  2011 November 10 were presented in 
\cite{Corbard2013}.
Figure \ref{Fig:assym} further shows an example of rotation-corrected $m$-averaged spectrum.
The first spatial leaks $\Delta\ell=\pm 3$ are visible with a clear 
leakage asymmetry. Such asymmetry was already mentioned by \cite{Korzennik1998} in 
the MDI intensity data but remained unexplained. However, \cite{HillandHowe1998} 
discussed that an error in the image radius can lead to a leakage asymmetry around 
the target mode. A proper understanding of all instrumental effects affecting the
geometry of SODISM or HMI images is required to properly build the leakage matrix.

 HMI continuum images at 607~nm covering the same 209-day period were first reduced to 
$256^2$ pixels and then processed through the same pipeline as SODISM images. 
HMI images are recorded every 45~s with a duty cycle of 98\% while SODISM images, recorded
every minute, reach only 74 \% of duty cycle. The signal-to-noise ratio in the power spectra 
is clearly better for HMI for which we were able to fit about 100 more modes with $l<100$
(865 modes versus 769 for SODISM).
 Up to 9 splitting coefficients were fitted for each mode and Fig.~\ref{Fig:comprot} shows a comparison 
of the rotation profiles obtained from both datasets by inverting the frequency splittings. 
In both cases we used a regularized least square method \citep{corbard1997} together with the
 generalized cross validation criteria of \cite{Wahba1977} for the choice of the regularization 
parameter. The uncertainties associated to each fitted splitting coefficient are propagated 
through the inverse process leading to the formal error bars 
shown on Fig.~\ref{Fig:comprot} (dotted lines). The results obtained from the two datasets remain
compatible within one sigma for all latitudes and depths. 
The main differences are found very close to the surface 
where higher $l$ modes would be needed to increase the resolution and at the equator where the 
local maximum found 
at $0.93 R_\odot$ from SODISM data \citep[see also][]{Corbard2013} is less pronounced with HMI.

\articlefigure[angle=-90,width=0.75\textwidth]{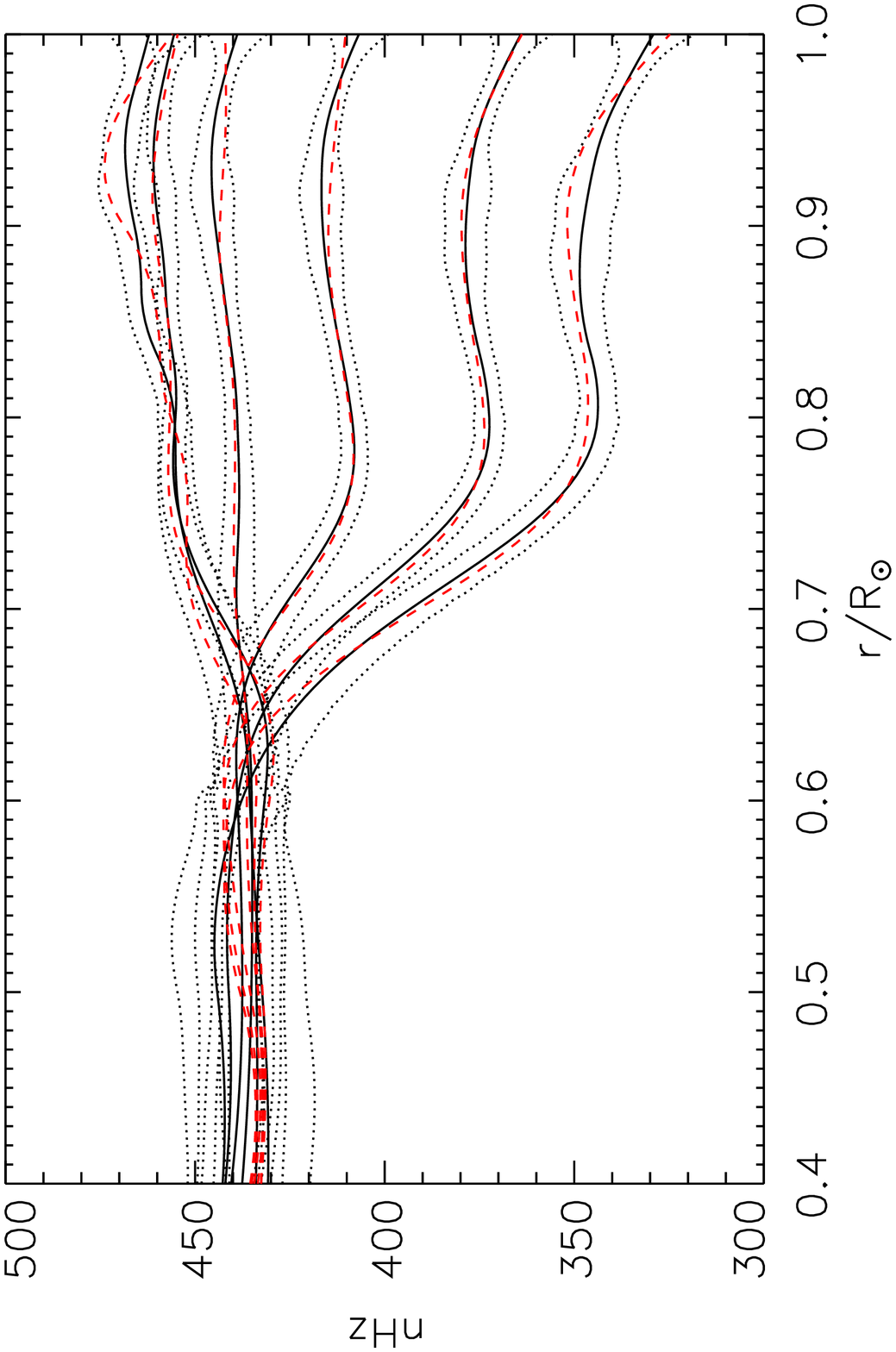}{Fig:comprot}{Internal rotation 
rate as a function of the fractional solar radius from the radiative 
interior up to the photosphere for latitudes of $0^\circ$, $15^\circ$, $30^\circ$, $45^\circ$, $60^\circ$ 
and $75^\circ$ from top to bottom.
 Results obtained from HMI intensity data (full lines) are shown with their one sigma formal error 
bars (dotted lines). SODISM results are shown by the dashed lines. }

\section{Conclusions}
Because of the low orbit of the satellite and the interruptions of the helioseismic sequence for 
other science objectives, the calibration of SODISM helioseismology data is a difficult task. 
The first comparison with HMI continuum intensity data shows however that we have reached a 
first satisfactory level of calibration. 
In order to fit higher $l$ modes, a major issue for both HMI and SODISM is to properly model 
the leakage matrix. Comparison between the two datasets can help for this
and we have now developed a full pipeline that is able to handle intensity images from 
both instruments.

\acknowledgements We acknowledge the work of SODISM and HMI instrument teams who provided the data. 
  D.~Salabert  acknowledges financial support from CNES.

\bibliography{TCorbard}

\end{document}